\documentstyle[12pt,aaspp4]{article}

\lefthead{Tanihata et al.}
\righthead{X--ray observations of BL Lac 97 July outburst}

\begin{document}

\title{Rapid Synchrotron Flares from BL Lacertae \\
	Detected by ASCA and RXTE}

\author{Chiharu Tanihata\altaffilmark{1,2} , 
	Tadayuki Takahashi\altaffilmark{1,2},
	Jun Kataoka\altaffilmark{1,2}, \\
	Greg M. Madejski\altaffilmark{3,4},
	Susumu Inoue\altaffilmark{5},
	Hidetoshi Kubo\altaffilmark{6}, \\
	Fumiyoshi Makino\altaffilmark{7},
	John R. Mattox\altaffilmark{8},
	and Nobuyuki Kawai\altaffilmark{9}}

\altaffiltext{1}{Institute of Space and Astronautical Science,
 	3-1-1 Yoshinodai, Sagamihara, 229-8510, Japan}
\altaffiltext{2}{Department of Physics, University of Tokyo,
	7-3-1 Hongo, Bunkyo-ku, Tokyo, 113-0033, Japan}
\altaffiltext{3}{Laboratory for High Energy Astrophysics, 
	Code 662, NASA/GSFC, Greenbelt, MD 20771, USA}
\altaffiltext{4}{Department of Astronomy,
	University of Maryland, College Park, MD 20742, USA}
\altaffiltext{5}{Theoretical Astrophysics Division,
	National Astronomical Observatory
	2-21-1 Ohsawa, Mitaka, Tokyo, 181-8588, Japan}
\altaffiltext{6}{Department of Physics, Tokyo Institute of Technology,
	Tokyo, 152-8551, Japan}
\altaffiltext{7}{Space Utilization Research Program, 
	Tsukuba Space Center, National Space Development Agency of Japan, 
	Tsukuba, Ibaraki 305-8505, Japan}
\altaffiltext{8}{Astronomy Department, Boston University, 
	Boston, MA 02215}
\altaffiltext{9}{The Institute of Physical and Chemical Research, 
	Wako, Saitama, 351-0198, Japan}

\begin{abstract}
We report the variable X--ray emission from BL Lacertae detected in  
the $ASCA$ ToO observation conducted during the $EGRET$ and $RXTE$
pointings, coincident with the 1997 July outburst.  The source showed 
a historically high state of X--ray, optical, and $\gamma$--ray emission, 
with its  2 -- 10 keV flux 
peaking at $\sim 3.3 \times 10 ^{-11}$ erg cm$^{-2}$ s$^{-1}$.
This is more than 3 times higher than the value measured by $ASCA$ in 1995.  
We detected two rapid flares which occured only in the 
soft X--ray band, while the hard X--ray flux also increased, 
but decayed with a much longer time scale.  
Together with the requirement of a very steep and varying 
power law dominating the soft X--ray band in addition to the hard
power law,
we suggest that both the high energy end of the
synchrotron spectrum and the hard inverse Compton spectrum were 
visible in this source during the outburst.
We discuss the possible origins of the observed variability
time scales, and 
interpret the short time scales of the soft X--ray variability 
as reflecting the size of the emission region.

\end{abstract}



\clearpage
\section{INTRODUCTION}

Rapid variability, together with the non-thermal emission observed
from radio to $\gamma$--rays, and very 
weak emission lines, are the main characteristics of BL Lac objects.
Those objects are a subclass of blazars, and the most promising 
explanation for these characteristics is the scenario where the 
radiation is emitted by energetic electrons contained in a
relativistic jet.  This jet points close to the line of sight, and
thus the radiation is beamed and Doppler-boosted (\cite{bla79}).  

Recently, many simultaneous multi-wavelength observations have been 
held, providing important tests for the emission mechanisms in blazars.
The broadband spectra of blazars consist of two peaks, 
one in the radio to optical--UV range (and in some cases, reaching to
the X--ray band), and the other in the X--ray to $\gamma$--ray region. 
From the high polarization of the radio to optical emission,
the lower energy peak is believed to be produced 
via the synchrotron process by relativistic electrons in the jet.
The higher energy peak is believed to be due to Compton upscattering 
by the same population of relativistic electrons.  Several
possibilities exist for the source of the seed photons;  these can 
be the synchrotron photons internal to the jet 
(e.g. \cite{jones74}; \cite{ghisellini89}), 
but also external, such as from the broad emission line clouds
(\cite{sikora94}) or from the accretion disk (\cite{dermer92};
\cite{dermer93}).
BL Lac objects are often separated into two groups 
by whether the synchrotron peak frequency is in the
UV--X--ray  region (HBLs, or High-Energy peaked BL Lacs) or IR--optical
region (LBLs, or Low-Energy peaked BL Lacs) 
(\cite{pad96}; \cite{sam96}; \cite{kubo98}).

BL Lacertae is the prototype object for the BL Lac class, 
and thus is one of the best studied blazars;  it is an LBL.  
The active nucleus lies in a giant elliptical galaxy at a 
redshift of 0.069 (\cite{miller78}).  While the optical and UV spectra of
BL Lac type objects are usually devoid of emission lines, 
the unusual aspect of of BL Lacertae 
is the relatively recent appearance of optical 
emission lines (\cite{vermeulen95}; \cite{corbett00}), 
obscuring the boundary between quasars and BL Lacs.  
Other than the emission lines, LBLs tend to have properties 
intermediate between quasars and HBLs such as luminosities 
or frequencies where the peaks are located 
(\cite{kubo98}; \cite{ghisellini98}; 
\cite{fossati98}).  The 1988 $Ginga$ observation of BL Lacertae 
showed that the X--ray spectrum of BL Lacertae is relatively flat, 
with a photon index of 1.7 to 2.2, and located above the extrapolation 
of the optical-UV continuum (\cite{kawai91}).
In 1995 it was observed with $ASCA$, showing the photon index 
$\Gamma$ = 1.94 $\pm$ 0.04 (\cite{greg99};  \cite{rita99});  it 
was also observed by $ROSAT$, which showed $\Gamma$ of 
1.94$^{+0.46}_{-0.47}$ (\cite{urry96}), and by $Einstein$, where the index was 
1.68 $\pm$ 0.18 (\cite{breg90}).  
In 1995, the $\gamma$--ray emission above 100 MeV 
was detected for the first time by the EGRET instrument 
onboard the Compton Gamma-Ray Observatory, or $CGRO$ (\cite{cat97}). 

BL Lacertae was observed by $ASCA$ following the report of 
a major outburst in 1997 discovered by optical observations 
(\cite{noble97}; \cite{mae97}).  Such outbursts are particularly 
valuable for investigating correlations between different wavelengths,
which led many observatories to point at BL Lacertae.  
EGRET observations indicated that the flux level above 100 MeV was 3.5 
times higher than that observed in 1995 (\cite{har97}; \cite{bloom97}).
It was also detected in the 50 -- 300 keV OSSE range (\cite{grove97}),
and 2 -- 10 keV $RXTE$ range (\cite{greg97}; \cite{greg99}).  
Intra-day time variability was observed in the optical band 
(\cite{mas98}; \cite{nesci98}) and in the X--ray band with 
$ASCA$ (\cite{maki97}).  
Fits to the spectral energy distributions have been done by 
several authors, and
interestingly, they all argue that the
high energy emission detected from BL Lacertae is likely to reveal 
Comptonization of {\sl both} internal (synchrotron) and external 
(broad-line) photons (\cite{greg99}; \cite{rita99}; \cite{bottcher00}).  

In this paper, we present and interpret the results of the $ASCA$ 
observation of BL Lacertae, together with the simultaneous $RXTE$ 
observations.  The observations are presented in Section 2.  In 
Section 3, we describe in detail the different variability patterns 
seen in the soft and hard X--ray bands;  in Section 4, we discuss
the radiative processes responsible for the X--ray emission, and 
summarize our results in Section 5.  

\section{OBSERVATIONS AND DATA REDUCTION}

We observed BL Lacertae with $ASCA$ during 1997 July 18.60 -- 19.62 UT.
The data were extracted by using the standard $ASCA$ procedures 
for SIS (Solid-state Imaging Spectrometer; \cite{burke91};
\cite{yamashita97}) and GIS (Gas Imaging Spectrometer; \cite{ohashi96}). 
GIS was used in the PH-nominal mode, and SIS was used in 1 CCD mode.
Source photons were extracted from a circular region centered at 
the target with 3$'$ and 6$'$ radii for SIS and GIS, 
respectively.  The background data were taken from source-free 
regions in the same respective images.  During the net exposure of 37 ks for 
SIS and 60 ks for GIS, the average count rates in the energy 
range of 0.7 -- 7.0 keV were 0.47 count s$^{-1}$, 0.41 count s$^{-1}$, 
0.33 count s$^{-1}$, 0.43 count s$^{-1}$, for SIS0, SIS1, GIS2, GIS3, 
respectively.  The $RXTE$ data were the same as presented in \cite{greg99}.  

During the 1997 July outburst, BL Lacertae was monitored intensively 
in the optical, and in the high energy $\gamma$--rays by EGRET.  Comparing 
the optical and $\gamma$--ray light curves measured over the 10-day 
span shown in Figure 2 of \cite{bloom97} to the $ASCA$ light curve shown in 
Figure \ref{fig:lc}, one can see that our 1-day observation occured 
exactly at the time of the correlated optical / $\gamma$--ray flare.  
In the the top and middle panels of Figure \ref{fig:lc}, we plot the 
soft (0.7 -- 1.5 keV) and hard (3 -- 7 keV) X--ray light curves.  The 
X--ray flare -- which occured close to the time of the optical and GeV 
$\gamma$--ray flares -- is strongly suggestive of a correlation of
emission in the three wavebands.
During the increase of the X--ray flux by a factor of nearly 4,
the flux increased by a factor of 2.5 in the EGRET band (30 MeV -- 30 GeV),
and a factor of 4 in the optical band (\cite{bloom97}).  However, we 
must note that the correlation (and any lags) cannot be precisely
measured because of the rather sparse sampling and limited sensitivity
of EGRET (but should be accomplished by the upcoming missions such
as $GLAST$).

Most importantly, we discovered an apparent difference between the 
soft and hard X--ray variability, and this is well illustrated in 
the bottom panel of Figure \ref{fig:lc}, showing the time series of 
the calculated hardness ratio.  In the soft X--ray band, we detected
two rapid flares with the rise and decay times on the order of 2 -- 3
hours.  The hard X--ray flux also flared up, but with a somewhat smaller
amplitude, and decreased with a longer time scale.  This suggests that 
two separate emission processes may be responsible for the X--ray 
emission from this source.  We also had two simultaneous $RXTE$ 
observations, labeled as [a] and [b] in Figure \ref{fig:lc}, which 
extended our energy range up to 20 keV.  

The average 2 -- 10 keV flux during the observation was $\sim 2.6 \times
10^{-11}$ erg cm$^{-2}$ s$^{-1}$, which is about 3 times higher than the
previous $ASCA$ observation in 1995 (\cite{greg99}; \cite{rita99}).
We fitted the $ASCA$ spectrum with a power law with absorption 
(model {\tt wabs + pegpwrlw} using XSPEC version 10.0).  We fixed the
absorption at the the Galactic value, 
$N_{\rm H} = 4.6\times 10^{21}$ atoms cm$^{-2}$, which is a sum of gas
columns measured in H and CO as summarized in \cite{greg99}.
Here the $ASCA$ data (using 0.7 -- 7.0 keV)
showed a significant excess below $\sim 1$ keV.
The excess is also indicated from the low value of absorption
when it was allowed to vary in the fitting process.  
On the other hand, 
the simultaneous $RXTE$ data (using 3-- 20 keV)
 are well fitted by the same model.
The fitting results are summarized in Table \ref{tbl:avrg}.  
We also found an indication of harder spectrum compared with
previous observations in 1995 by $ASCA$ 
($\Gamma$ = 1.94 $\pm$ 0.04; \cite{greg99})
and in 1988 by $Ginga$ 
($\Gamma$ = 1.7 $\pm$ 0.1, 2.2 $\pm$ 0.2; \cite{kawai91}).



\section{THE TWO - COMPONENT NATURE AND VARIABILITY OF THE X--RAY SPECTRUM}

Because the source was varying, we first selected two sets of 
spectra including the two pairs of $RXTE$ and $ASCA$ data sets, 
labeled as [a] and [b] in Figure \ref{fig:lc}.  Note that one 
set was during a state of relatively low flux (which is still 
higher than the 1995 value), and the other close to the maximum 
of the flaring state;  this provided two different flux states 
for a comparison of the spectra.  As reported in \cite{greg99}, 
spectral variability (although at a modest level) was seen 
among various segments of the $RXTE$ observation.  

Before performing joint spectral fits to the $ASCA$ and $RXTE$ data, 
we selected the common energy range of 3 -- 7 keV for both data sets 
to verify the consistency between the two instruments.  In selecting
the same time regions for $ASCA$, we extended the time intervals 
to 5 ks in order to collect a sufficient number of counts.  We 
fitted both $ASCA$ and $RXTE$ data with a single power law plus 
Galactic absorption model.  For all the spectral fitting hereafter, 
we combined the data from 2 SISs and 2 GISs for the $ASCA$ data, and 
the data from PCU 0, 1, and 2 for the PCA data.  The resulting best
fits for region [a] and [b] for $ASCA$ were:  photon index 
$\Gamma = 1.5 \pm 0.4$ and $1.4 \pm 0.2$, with 3 -- 7 keV flux 
$F_{\rm 3-7keV}$ = 1.1 $\pm$ 0.1 and 1.67 $\pm$ 0.09 $\times$ 
10$^{-11}$ erg cm$^{-2}$ s$^{-1}$.  For the $RXTE$ data, they 
were:  $\Gamma$ = 1.60 $\pm$ 0.15 and 1.4 $\pm$ 0.1, $F_{\rm 3-7keV}$ 
= 1.28 $\pm$ 0.04 and 1.95 $\pm$ 0.04 $\times$ 10$^{-11}$ erg cm$^{-2}$ 
s$^{-1}$.  This indicates that the photon indices are consistent, 
but the flux inferred by the $RXTE$ is larger by $\sim$10\%.
Regarding this as an uncertainty of the calibration of two
instruments, we added a multiplying parameter ({\tt constant} 
in XSPEC) to account for the difference in the normalization. 
We fixed the value to 1 for $ASCA$, and allowed it to vary for $RXTE$.
We fitted the 3 -- 7 keV spectrum of  $ASCA$ and $RXTE$ together,
and resulted with a normalization factor of $RXTE$ to be
1.1 $\pm$ 0.1 and 1.16 $\pm$ 0.06 for regions [a] and [b].  

We then fitted the full 0.7 -- 20 keV combined data set
with a single power law plus Galactic absorption model,
with the normalization of the PCA data kept free.  We found that although the 
single power law model provides an adequate fit to the low state 
spectrum [a], it is rejected for the flaring state spectrum [b] 
at a confidence level of more than 99.9\%.  Thus we fitted the spectrum 
[b] with an additional soft power law, where the fit improved to be 
acceptable. Again, the normalization factors of $RXTE$ were 
1.08 $\pm$ 0.06 and 1.14 $\pm$ 0.05 for regions [a] and [b], 
which is within the error bars of the values derived above.
The fitting results to the combined data 
are summarized in Table \ref{tbl:sum}.

The unfolded spectra of the two time regions [a] and [b] are shown in
Figure \ref{fig:spec}.  It is apparent from that Figure that a 
power law with a hard index is dominant in both spectra, which is 
consistent with the fact that BL Lacertae is classified as an LBL;    
in the context of the scenario described in the Introduction, 
the main emission process in the X--ray band above $\sim 1$ keV is 
the inverse Compton process.  However, for the flaring 
state spectrum, we found that another soft spectral component, 
well-described as a steep power law, begins to dominate below $\sim 1$ keV.  

Any detailed modeling of the source requires the knowledge of fluxes
of the two spectral components {\sl separately}.  Since both the $E < 1$
keV and $E > 1$ keV count rates are affected by both components and
not just by the soft or hard component alone, we had to model the 
time-resolved spectrum to unravel the light curves of each component.  
This is particularly important since the combined $ASCA$ and $RXTE$ 
data revealed the soft component varying with larger amplitude at 
flaring states.  We separated the entire observation into 9 time regions, 
labeled as (1) -- (9) in the top panel of Figure \ref{fig:lc}, and
fitted the entire spectrum with a model including two power laws plus 
Galactic absorption.  Since we found that the $ASCA$ spectrum above 
2 keV can be well fitted by a single power law plus Galactic absorption 
model, we inferred that the spectrum above 2 keV is only modestly 
affected by the soft component.  Thus, we first fitted the $ASCA$ 
data set above 2.5 keV with a single power law plus Galactic 
absorption to determine the initial parameters of the hard power law.
In the subsequent fitting  of the full spectrum,
we kept all four parameters of the two power laws free.
All such spectra were well fitted by the two power law plus Galactic 
absorption model.  The index and the flux of each component are 
plotted in Figure \ref{fig:index}.  We found that the soft power law 
is very steep with the photon index $\sim$ 3 -- 5 which appears to 
vary, while the hard power law has an index that varies little, 
between $\sim 1.3 - 1.6$.  The flux variability is a factor 
of 2 for the hard component, and a factor of 4 for the soft component.  
The derived fitting parameters are summarized in Table \ref{tbl:each}.

\section{DISCUSSION}

The $ASCA$ observation of BL Lacertae showed that the flux is highly
variable on a time scale of hours, and that the variability in the 
soft and hard X--ray bands appears significantly different, indicating
that two different emission processes are responsible for the X--rays
detected by $ASCA$.  Perhaps the most natural explanation of this
behavior is where the high energy ``tail'' of the rapidly flaring 
synchrotron component is mixing into the less variable 
inverse Compton spectrum.  The very steep power law required in the
soft spectrum in addition to the hard power law also supports this idea.
The existence of the soft component was also suggested from the 
spectrum in the previous $ASCA$ observation in 1995 (\cite{rita99}; 
\cite{greg99}).  However, the reality of this soft component as 
intrinsic to BL Lacertae -- inferred on the basis of the excess of
soft X--ray flux -- was somewhat uncertain, as it also could be 
interpreted by the overestimated value of the CO absorption.  
Our $ASCA$ results of 1997 July are the first to confirm the 
existence of the soft component in BL Lacertae not solely based on 
the photon spectrum, but on the different variability patterns of both
components.   This will be the third source following S5 0716+714 
(\cite{cap94}; \cite{giommi99}) and ON~231 (\cite{taglia00})
which reveals two components inferred
on the basis of the difference in the variability patterns. 
We thus suggest that the soft power law dominating the soft X--ray
band is the high energy tail of the synchrotron spectrum, representing 
its steep cutoff.  Because the synchrotron emission frequency is 
proportional to the square of the energy of the radiating electrons, 
we can regard the 
soft X--ray band as reflecting the most energetic electrons. 
The flat index of the hard component indicates that
the hard X--rays are representing the onset of the inverse Compton 
spectrum reflecting the lower energy end of the electron population, 
consistent with what we have known from previous observations.

The overall spectral energy distribution (SED) of BL Lacertae 
during the July 1997 outburst is shown in Figure \ref{fig:multi};
this Figure also illustrates the literature and archival data for 
pre-1997 observations.  The two peaks common in blazar spectra are clearly 
seen, with the higher energy part of our combined $ASCA$ and 
$RXTE$ spectrum smoothly connecting to the $\gamma$--ray band 
observed by OSSE and EGRET.  The upturn in the soft X--rays is evident, 
illustrating the synchrotron spectrum extending to the $ASCA$ band.
The X--ray light curve of BL Lacertae shown in Figure \ref{fig:lc}
reveals that the rapid flares are more prominent in the soft X--ray 
band.  One possibility is that the flare is caused by the increase of the 
maximum energy of the electron population, similar to the flares 
observed in several HBLs, such as from Mrk~501 
(\cite{kataoka99}; \cite{pian98}).
However, we must remark that BL Lacertae also requires some other
mechanism to account for the changes in the hard X--ray flux, which
presumably forms the low energy end of the SSC spectrum.  The most 
likely explanation for this is the increase of the normalization 
of the electron population, although variability of the magnetic 
field or the beaming factor is also possible.  

Rapid variability in blazars is generally assumed to be triggered by 
shocks propagating along a relativistic jet, and such a shock front 
is thought to  be the most promising region where the electron 
acceleration may be taking place.  The cutoff (the maximum) 
of the electron population is likely to result from the electron 
cooling time scale $\tau_{\rm cool}$ balancing the acceleration time scale 
$\tau_{\rm acc}$ in this region (e.g. \cite{inoue96}; \cite{kirk98}; 
\cite{kusunose00}).  In this case, the photon spectrum also cuts off 
at a corresponding frequency, 
$\nu_{\rm max} \sim 1.2 \times 10^6 \,B \,\delta \,\gamma_{\rm max}^2$
(using the peak frequency of the synchrotron spectrum; see, 
e.g., \cite{rybicki}), followed by a steeper decline.
Here, $B$ is the magnetic field in units of Gauss, 
and $\gamma_{\rm max}$ is the Lorentz factor of the
maximum energy electrons.
Because the
photon index of the soft component in our $ASCA$ data was mainly
steeper than 3, the cutoff of the synchrotron spectrum most likely 
occurs at an energy lower than the $ASCA$ band, $\sim$ 0.1 keV or less.  
Using 0.1 keV, 
we can calculate, 
$\gamma\/_{\rm max} \leq 4.5 \times 10^4 \,(h\nu/{\rm 0.1keV})^{0.5} \, 
B_1^{-0.5} \, \delta_{10}^{-0.5}$, 
where $B_1$ is the magnetic field in units of Gauss, and $\delta_{10}$ 
the beaming factor $\delta / 10$.  We used $B = 1$ G and $\delta = 10$ 
inferred from the fits to the SED by \cite{greg99} and \cite{rita99}.  

From the calculated maximum energy of the relativistic electrons, 
we can now estimate the energy at which the overall spectrum cuts 
off.  As suggested by \cite{greg99} and \cite{rita99}, assuming 
that the highest energy ($\gamma$--ray) range of the spectrum is 
produced by the Compton mechanism via scattering of external 
photons from the broad line region, the overall spectrum should extend
only up to the energy corresponding to these photons interacting with 
the most energetic electrons.  
If Klein-Nishina effects become important,
the spectrum should also steepen above a certain energy.
Taking the typical energy of the broad line photons to be
$h \nu_{\rm BL} \sim$ 1.9 eV (or wavelength $\sim$ 6600 \AA,
the H$\alpha$ line) and $\delta$ = 10, 
Klein-Nishina effects set in for electrons with 
Lorentz factors exceeding 
$\gamma_{\rm KN} \sim m_e c^2 / (\delta h \nu_{\rm BL}) 
\sim 2.7 \times 10^4$, 
where $m_e$ is the electron mass, and $c$ the velocity of light.
This is slightly below the maximum Lorentz factor inferred from
the soft X--ray cutoff.
Thus the $\gamma$--ray spectrum should break above 
$h \nu_{\rm KN} 
\sim \delta^2 h \nu_{\rm BL} \times \gamma_{\rm KN}^2
\sim 140$ GeV.
The maximum attainable $\gamma$--ray energy corresponds to the
maximum electron energy, 
$h \nu_{\rm max} \sim \delta \gamma_{\rm max}^2 m_e c^2 
\sim 230$ GeV.
Accordingly, BL Lacertae is not likely to be a TeV emitter, 
consistent with the observations showing only upper limits (\cite{cat97}).  
This is in contrast to the ``TeV blazars'' such as Mrk~421 or Mrk~501,
where the $\gamma_{\rm max}$ is thought to extend up to $\sim10^6$, and 
this high value of the maximum electron energy should be one of the main 
factors responsible for a blazar being a TeV emitter.  

The decay time of the variability in the $ASCA$ data for BL Lacertae
appears to be longer for hard X--rays -- presumably produced by
the low end of the distribution of electron energies -- than for 
soft X--rays, produced by the most energetic electrons.  This is 
reasonable given the fact that the electron cooling time is shorter 
for higher energy electrons 
($\tau_{\rm cool} \simeq 7.7 \times 10^8 \, (1 + u_{\rm soft} / 
u_{\rm B})^{-1} \, B^{-2} \, \gamma^{-1}$ (see, e.g. \cite{rybicki}), 
where $u_{\rm B}$ and $u_{\rm soft}$ are the energy density of the 
magnetic field, and the soft photons to be upscattered.  
As the scattering is expected to occur mainly in the 
Thomson regime, $u_{\rm soft}/u_{\rm B}$ can be estimated from the 
peak luminosity of the synchrotron ($L_{\rm sy}$) and inverse Compton 
($L_{\rm IC}$) components in the observed spectrum
(cf. Fig.~\ref{fig:multi}) via 
$u_{\rm soft}/u_{\rm B} = L_{\rm IC}/L_{\rm sy} \sim 4$.  
Applying $B\sim 1$ Gauss and $\delta \sim 10$, the cooling time 
for the maximum energy electrons is calculated to be 
$\tau_{\rm cool,obs}(\gamma_{\rm max}) = \tau_{\rm cool,intr}/\delta
\leq 3.5 \times 10^2 \,(h\nu/{\rm 0.1keV})^{0.5} \, B_1^{-1.5} \, 
\delta_{10}^{-0.5}$ s,  
which is much shorter than the X--ray variability
time scales of 2 -- 3 hours that we have observed.  
This can be interpreted as resulting from the fact that
time scales faster than the source 
light-crossing time $\tau_{\rm crs}$, which is the time for the photons 
to propagate across the emission region, will always be smoothed out 
by $\tau_{\rm crs}$ (cf. \cite{chia99}).  

The soft X--ray light curve indicates that the rise and decay times are 
similar to each other.  This is nicely consistent with the scenario
where the soft 
X--ray band reflects the highest energy electrons, both 
$\tau_{\rm cool}$ and $\tau_{\rm acc}$ are shorter 
than the $\tau_{\rm crs}$, and
thus both rise and decay times are reflecting the $\tau_{\rm crs}$.
Regarding the {\sl hard} X--rays, in the context of the two-component 
Comptonization model (cf. \cite{greg99}), those are produced by lower energy
electrons, where the seed photons for the Comptonized spectrum
are probably the synchrotron photons in the optical band, 
at $\nu \sim 10^{14}$ Hz.  With this, the 5~keV ($\sim 10^{18}$ Hz) 
X--rays are produced by electrons with $\gamma \sim 100$, and for those
electrons, $\tau_{\rm cool}$ is about 40 hours.  This is now significantly
longer than the $\tau_{\rm crs}$, estimated to be 2 -- 3 hours,  
so those electrons are not expected to cool within $\tau_{\rm crs}$.  
The long, gradual decay of the hard X--rays 
is then probably due to other causes, such as escape of particles from the 
radiating region, or decrease in the effective density of soft
(target) photons.  

In contrast to the hard X--rays, adopting the above parameters -- and 
assuming that the region producing the optical radiation is co-spatial with
that producing the soft X--rays -- we infer that the cooling times of
the electrons producing the {\sl optical} synchrotron photons (which
should have $\gamma \sim 6500$ for $\lambda = 6000$ \AA) 
are $\sim 2500$ s, which, just as for the 
case of soft X--rays, is shorter than the source crossing time.  
This is supported by the very rapid variability of the optical 
flux with similar rise and decay times (\cite{mas98};
\cite{nesci98}), similar to that seen in the soft X--ray band.  

However, we must also remark that we cannot completely exclude the 
possibility of variability on a shorter time scale which we cannot
measure.  The detected flares can actually consist of a superposition 
of more rapid undetectable flares, and there is actually a data point 
indicating a more rapid variability in the middle of the second flare, 
but the statistics do not let us distinguish such small flares.  
We expect future missions with larger effective area to provide
sufficiently sensitive observations to completely resolve the most
rapid events.  

\section{CONCLUSIONS}

$ASCA$ X--ray observations of BL Lacertae during the July 1997 outburst
revealed that the X--ray flux was in a high state throughout the
observation, in similarity to the optical and high energy
$\gamma$--ray behavior.  We detected two rapid flares with a time 
scale of 2 -- 3 hours, but 
only in the soft X--ray band.  The hard X--ray flux also increased, but
decayed with a much longer time scale.  The spectrum of the source in
the highest flux states required two separate power law components,
a soft and steep one, dominating below $\sim 1$ keV with a photon 
index of 3 -- 5, in addition to a hard component with a photon index
1.3 -- 1.6.  This suggests that the soft component represents the high 
energy end of the synchrotron spectrum extending to the X--ray band,
and thus allows us to estimate the position of the high energy 
cutoff of the synchrotron component, from which we
infer the cutoff in the electron energy distribution to be 
$\gamma_{\rm max} \sim 4.5  \times 10^{4}$.  The cooling time scale of
such electrons is $\sim$ 350 s, short as compared to the variability
time scale $\tau_{\rm var}$.   We interpret this $\tau_{\rm var}$ as the
indicator of the source size.  The slower decay time of the hard
X--ray flux fits well into the scenario where the hard
X--rays are produced via the Comptonization of optical synchrotron photons
by the lower energy end of the relativistic electron distribution (where the
cooling time scales are longer than the source crossing time).  
In this context, the long decay time in the hard X--ray flux is due 
to other causes, such as escape of particles from the radiating region,
or decrease in the effective energy density of soft photons.

\acknowledgments
This research was supported by the Fellowship of the
Japan Society for Promotion of Science for Young Scientists,
Grant-in-Aid for Scientific Research No. 05242101,
the Grand-in-Aid for COE Research No. 07CE2002 by the
Ministry of the Education, Culture and Science, Japan, and NASA
grants to USRA and University of Maryland / College Park, 
and also by NASA Grant NAG5-3384 to Boston University.



%
%

\figcaption[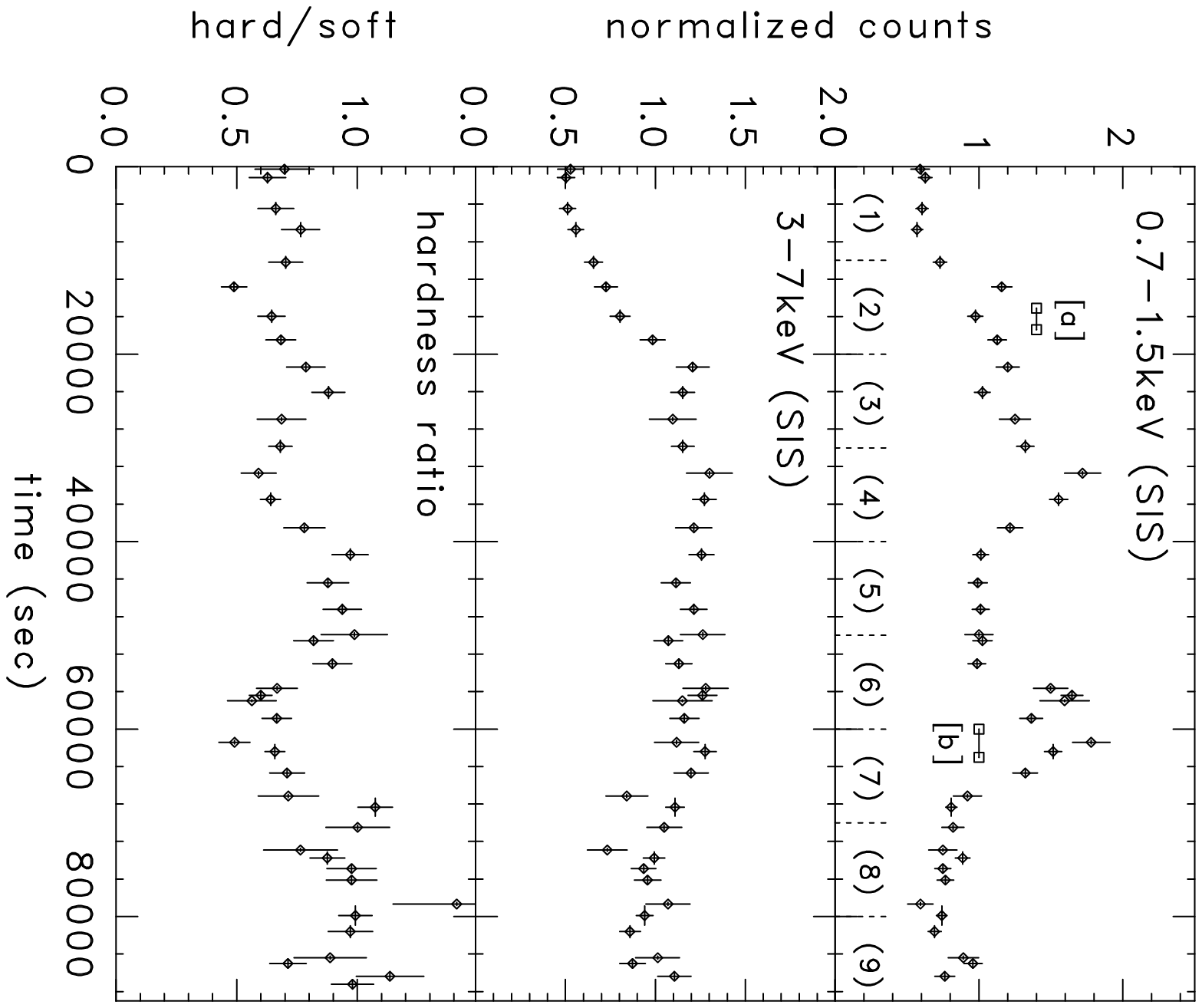]{X--ray light curve of BL Lacertae 
        during 1997 July 18.6 -- 19.6; 
	top: $ASCA$ soft band (0.7 -- 1.5 keV);
        middle: $ASCA$ hard band (3 -- 7 keV);
        bottom: hardness ratio, defined as hard X--ray count rate / 
	soft X--ray count rate.
	Both top and middle panel rates are normalized by their
        respective average count rate during the observation.  \label{fig:lc}}

\figcaption[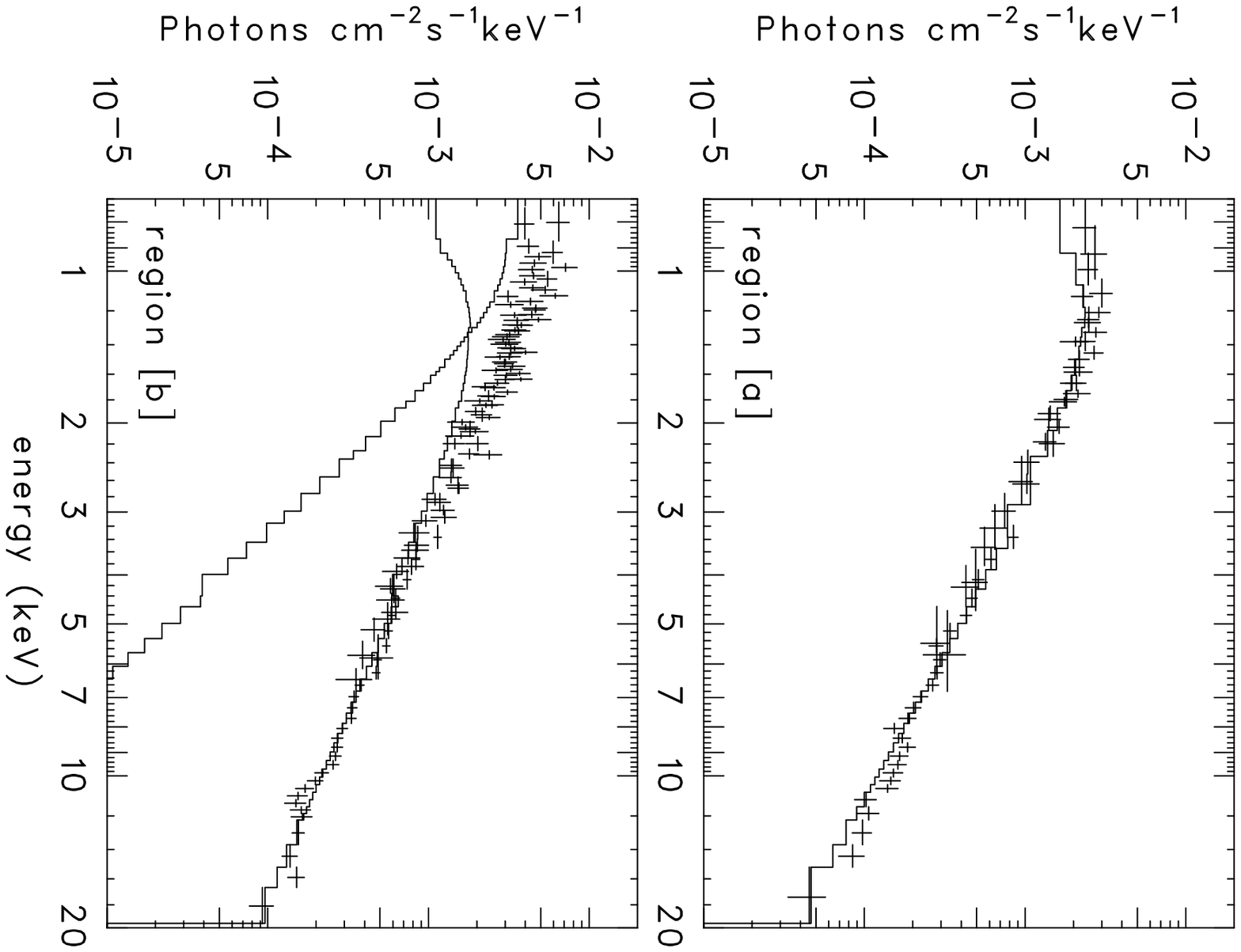]{Unfolded spectrum of BL Lacertae from 
	combined $RXTE$ and $ASCA$ data for the
	low state [a] and flare state [b] in Figure \ref{fig:lc}.
	The soft power law appears to dominate the lower energy X--ray
	band during the flare state. \label{fig:spec}}

\figcaption[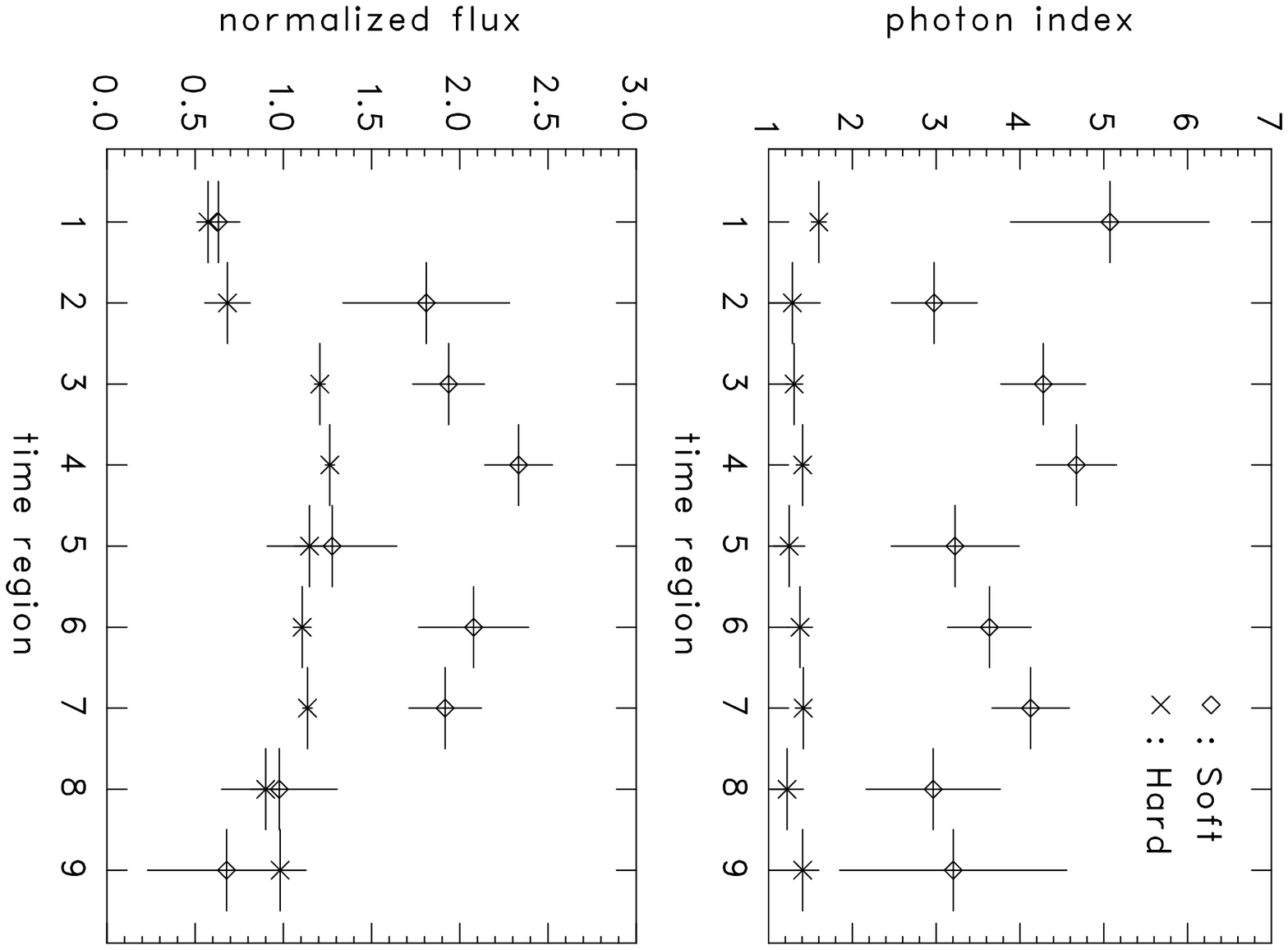]{Variability  of the photon
	indices and fluxes for both the soft and hard power law components.
	The absorption was fixed to the Galactic value, and
	the fluxes are in the 0.7 -- 1.5 keV band for the soft X--rays,
	and 3 -- 7 keV band for the hard X--rays, both
	normalized to the value at the first period. \label{fig:index}}

\figcaption[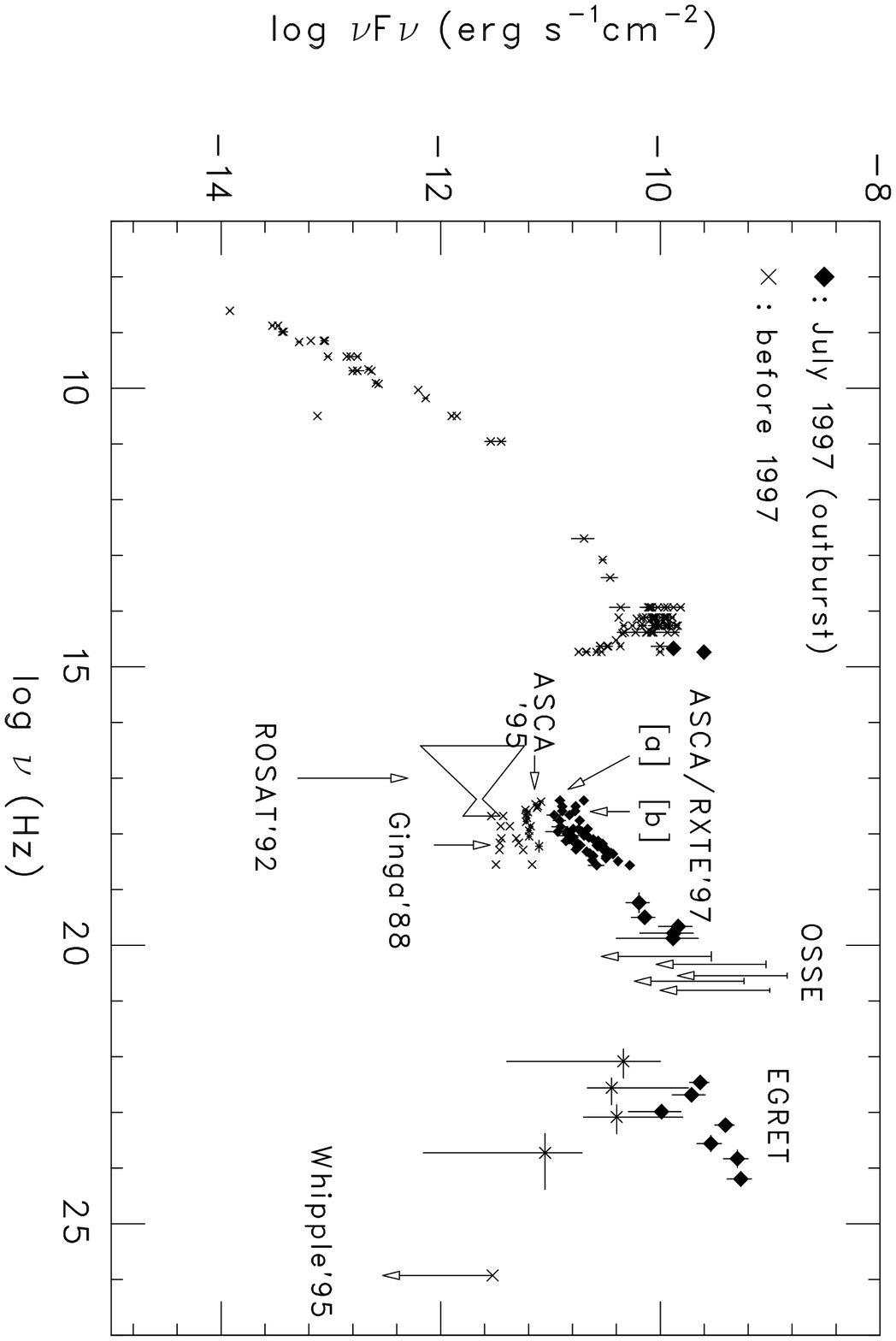]{Multi-frequency SED 
	of BL Lacertae during the July 1997 outburst,
	plotted together with the archival data.
	The diamonds represent the data during the outburst,
	and the two combined $ASCA$ and $RXTE$ data sets 
	correspond to the two time regions marked 
	as [a] and [b] in Figure \ref{fig:lc}.
	The $Ginga$ data are from Kawai et al.\ 1991;  
	the $ROSAT$ data are from Urry et al.\ 1996;  
	the EGRET data are from Bloom et al.\ 1997 and Catanese et
	al. 1997, and the Whipple data are from Catanese et al.\ 1997.   
	The OSSE and ASCA 1995 data is from the HEASARC public data,
	and the radio to optical data are from the NED data base. 
\label{fig:multi}}

\clearpage

\begin{deluxetable}{cccccc}
\tablecaption{Results of Spectral Fits to the $ASCA$ and $RXTE$ PCA 
	Data for BL Lacertae
	\label{tbl:avrg}}
\tablewidth{0pt}
\scriptsize
\tablehead{
\colhead{} & \colhead{Date (UT)} & \colhead{N$_{\rm H}$} &
	\colhead{} & \colhead{F$_{\rm 2-10keV}$} & \colhead{} \\
\colhead{Instrument} 
	& \colhead{July 1997} & \colhead{($\times$10$^{21}$ cm$^2$)} 
	& \colhead{$\Gamma$}
	& \colhead{($\times$10$^{-11}$erg cm$^{-2}$s$^{-1}$)}
	& \colhead{$\chi^2_{\nu}$ (dof)} 
}
\startdata
SIS & 18.60 -- 19.62 & 4.6\tablenotemark{a} &  1.75$\pm$0.01 &  2.52$\pm$0.03 & 1.93 (313)\nl
    & 		& 2.7$\pm$0.1 		&  1.51$\pm$0.02 &  2.71$\pm$0.03 & 1.27 (312)\nl
GIS & 18.60 -- 19.62 & 4.6\tablenotemark{a} &  1.72$\pm$0.01 &  2.61$\pm$0.02 & 2.98 (373)\nl
    & 		& 0.9$\pm$0.1 		&  1.39$\pm$0.02 &  2.78$\pm$0.03 & 1.84 (372)\nl
\tableline
PCA & 18.78 -- 18.80 ([a]) & 4.6\tablenotemark{a} &  1.58$\pm$0.06 &  2.50$\pm$0.01 & 0.72 (42)\nl
PCA & 19.30 -- 19.32 ([b]) & 4.6\tablenotemark{a} &  1.44$\pm$0.03 &  3.72$\pm$0.01 & 0.85 (43)\nl
\enddata
\tablenotetext{}{Note.-- All errors are 1 $\sigma$.}
\tablenotetext{a}{fixed to Galactic absorption}
\end{deluxetable}

\begin{deluxetable}{cccccc}
\tablecaption{Results of Spectral Fits to joint $ASCA$ and $RXTE$ PCA
Data for BL Lacertae
	\label{tbl:sum}}
\tablewidth{0pt}
\scriptsize
\tablehead{
\colhead{Model} & 
\colhead{$\Gamma_{\rm 1}$\tablenotemark{a}}  & 
\colhead{F$_{\rm 1}$\tablenotemark{b}}  & 
\colhead{$\Gamma_{\rm 2}$\tablenotemark{a}}  & 
\colhead{F$_{\rm 2}$\tablenotemark{b}}  & 
\colhead{$\chi^2_{\nu}$ (dof)} \\
\tableline
\multicolumn{6}{c}{Low State (Region [a]\tablenotemark{c}\ )}
}
\startdata
single power law &  1.76$\pm$0.04 & 2.4$\pm$0.1   & \nodata      & \nodata & 0.71 (100) \nl 
\cutinhead{Flare State (Region [b]\tablenotemark{c}\ )}
single power law &  1.68$\pm$0.03 & 3.6$\pm$0.1   & \nodata      & \nodata & 1.50 (166) \nl 
double power law &  3.7$\pm$0.4   & 0.23$\pm$0.01 & 1.38$\pm$0.06 & 3.1$\pm$0.2 & 0.65 (164) \nl 
\enddata
\tablenotetext{}{Note.-- The absorption is fixed to the Galactic value.}
\tablenotetext{}{Note.-- All errors are 1$\sigma$.}
\tablenotetext{a}{photon index}
\tablenotetext{b}{2 -- 10 keV flux in units of 10$^{-11}$ erg cm$^{-2}$ s$^{-1}$}
\tablenotetext{c}{the ASCA time region is extended
	to 5 ks in order to collect sufficient photons}
\end{deluxetable}

\begin{deluxetable}{cccccc}
\tablecaption{Results of Time-Resolved Spectral Fits to $ASCA$ 
Data for BL Lacertae
	\label{tbl:each}}
\tablewidth{0pt}
\footnotesize
\tablehead{
\colhead{Time Region} & 
\colhead{$\Gamma_{\rm soft}$\tablenotemark{a}} & 
\colhead{F$_{\rm soft}$\tablenotemark{b}} &
\colhead{$\Gamma_{\rm hard}$\tablenotemark{a}} & 
\colhead{F$_{\rm hard}$\tablenotemark{c}} &
\colhead{$\chi^2_{\nu}$ (dof)}
}
\startdata
1 & 5.1$\pm$1.2 & 2.4$\pm$0.5 & 1.6$\pm$0.1 & 7.7$\pm$0.2 & 1.04(223) \nl
2 & 3.0$\pm$0.5 & 6.9$\pm$1.8 & 1.3$\pm$0.3 & 9.1$\pm$1.7 & 0.96(254) \nl
3 & 4.3$\pm$0.5 & 7.4$\pm$0.8 & 1.3$\pm$0.1 & 16.1$\pm$0.4 & 0.89(245) \nl
4 & 4.7$\pm$0.5 & 8.9$\pm$0.7 & 1.4$\pm$0.1 & 16.9$\pm$0.4 & 1.14(314) \nl
5 & 3.2$\pm$0.8 & 4.8$\pm$1.4 & 1.2$\pm$0.2 & 15.4$\pm$1.2 & 0.87(288) \nl
6 & 3.6$\pm$0.5 & 7.9$\pm$1.2 & 1.4$\pm$0.1 & 14.8$\pm$0.7 & 0.93(321) \nl
7 & 4.1$\pm$0.5 & 7.3$\pm$0.8 & 1.4$\pm$0.1 & 15.2$\pm$0.4 & 1.14(353) \nl
8 & 3.0$\pm$0.8 & 3.7$\pm$1.2 & 1.2$\pm$0.2 & 12.0$\pm$1.1 & 1.14(322) \nl
9 & 3.2$\pm$1.4 & 2.6$\pm$1.7 & 1.4$\pm$0.2 & 13.1$\pm$1.2 & 1.04(297) \nl
\enddata
\tablenotetext{}{Note.-- 2 SISs and 2 GISs are 
		combined for the Spectral Fits; All errors are 1 $\sigma$.}
\tablenotetext{a}{photon index}
\tablenotetext{b}{0.7 -- 1.5 keV flux in units of 10$^{-12}$erg cm$^{-2}$s$^{-1}$}
\tablenotetext{c}{3.0 -- 7.0 keV flux in units of 10$^{-11}$erg cm$^{-2}$s$^{-1}$}
\end{deluxetable}

\clearpage
\begin{figure}
\epsscale{0.8}
\plotone{f1.ps}
\end{figure}
\clearpage
\begin{figure}
\epsscale{0.8}
\plotone{f2.ps}
\end{figure}
\clearpage
\begin{figure}
\epsscale{0.8}
\plotone{f3.ps}
\end{figure}
\begin{figure}
\epsscale{0.8}
\plotone{f4.ps}
\end{figure}
\vfill\eject


\clearpage

\begin{thebibliography}{}
\bibitem[Blandford \& Konigl 1979]{bla79} 
	Blandford, R. D., \& Konigl, A. 1979, \apj, 232, 34
\bibitem[Bloom et al.\ 1997]{bloom97} Bloom, S. D., et al. 1997, \apj, 490, L145
\bibitem[Bottcher \& Bloom 2000]{bottcher00}
        Bottcher, M., \& Bloom, S. D. 2000, \apj, 119, 469
\bibitem[Bregman et al.\ 1990]{breg90} Bregman, J. N., et al. 1990, \apj, 352, 574
\bibitem[Burke et al.\ 1991]{burke91} Burke, B. E., et al. 1991, 
  		IEEE Trans. ED-38, 1069 
\bibitem[Cappi et al.\ 1994]{cap94} Cappi, M., Comastri, A., Molendi, S., 
Palumbo, G. C. C., della Ceca, R., \& Maccacaro, T. 1994, \mnras, 271, 438 
\bibitem[Catanese et al.\ 1997]{cat97} 
	Catanese, M., et al.\ 1997, \apj, 480, 562
\bibitem[Chiaberge \& Ghisellini 1999]{chia99}
	Chiaberge, M., \& Ghisellini, G. 1999, \mnras, 306, 551
\bibitem[Corbett et al. 2000]{corbett00}
        Corbett, E. A., Robinson, A., Axon, D. J., \& Hough, J. H.
        2000, \mnras, 311, 485
\bibitem[Dermer, Schlickeiser, \& Mastichiadis 1992]{dermer92}
        Dermer, C. D., Schlickeiser, R., 
	\& Mastichiadis, A. 1992, \aap, 256, L27
\bibitem[Dermer \& Schlickeiser 1993]{dermer93}
        Dermer, C. D., \& Schlickeiser, R. 1993, \apj, 416, 458
\bibitem[Fossati et al. 1998]{fossati98}
        Fossati, G., Maraschi, L., Celotti, A., Comastri, A.,
        \& Ghisellini, G. 1998, \mnras, 299,433
\bibitem[Giommi et al.\ 1999]{giommi99} Giommi, P., et al. 1999, \aap, in press
\bibitem[Ghisellini \& Maraschi 1989]{ghisellini89}
	Ghisellini, G., \& Maraschi, L. 1989, \apj, 340, 181
\bibitem[Ghisellini et al.\ 1998]{ghisellini98}
	Ghisellini, G., Celloti, A., Fossati, G., Maraschi, L.,
	\& Comastri, A. 1998, \mnras, 301, 451
\bibitem[Grove \& Johnson 1997]{grove97} 
	Grove, J. E., \& Johnson, W. N. 1997, \iaucirc ~6705
\bibitem[Hartman et al.\ 1997]{har97} Hartman, R., Bertsch, D.,
	 Bloom, S., Sreekumar, P.,
	 Thompson, D., Ma, F., \& Barry, D. 1997, \iaucirc ~6703
\bibitem[Inoue \& Takahara 1996]{inoue96}
	 Inoue, S., \& Takahara, F. 1996, \apj, 463, 555
\bibitem[Jones, O'Dell, \& Stein 1974]{jones74} 
	Jones, T. W., O'Dell, S. L., \& Stein, W. A. 1974, \apj, 188, 353
\bibitem[Kataoka et al.\ 1999]{kataoka99} 
	Kataoka, J., et al. 1999, \apj, 514, 138
\bibitem[Kawai et al.\ 1991]{kawai91} Kawai, N., et al. 1991, \apj, 439, 80
\bibitem[Kirk, Rieger, \& Mastichiadis 1998]{kirk98}
	Kirk, J. G., Rieger, F. M., \& Mastichiadis, A. 1998, \apj, 333, 452
\bibitem[Kubo et al.\ 1998]{kubo98} Kubo, H., Takahashi, T.,
	 Madejski, G., Tashiro, M.,
	 Makino, F., Inoue, S., \& Takahara, F. 1998, \apj, 504, 693
\bibitem[Kusunose, Takahara, \& Li 2000]{kusunose00}
        Kusunose, M., Takahara, F., \& Li, H. 2000, \apj, in press
\bibitem[Madejski et al.\ 1997]{greg97} Madejski, G., Jaffe, T., 
	\& Sikora, M. 1997, \iaucirc ~6705
\bibitem[Madejski et al.\ 1999]{greg99} Madejski, G. M., Sikora, M.,
 	Jaffe, T., B\l a\.{z}ejowski, M., Jahoda, K., 
	\& Moderski, R. 1999, \apj, 521, 145
\bibitem[Maesano et al.\ 1997]{mae97} Maesano, M., Massaro, E., 
	\& Nesci, R. 1997, \iaucirc ~6700
\bibitem[Makino et al.\ 1997]{maki97} Makino, F., Mattox, J.,
	 Takahashi, T., Kataoka, J.,
	 \& Kubo, H. 1997, \iaucirc ~6708
\bibitem[Massaro et al.\ 1998]{mas98} Massaro, E., Nesci, R.,
	 Maesano, M., Montagni, F.,
	 \& D'Alessio, F. 1998, MNRAS, 299, 47
\bibitem[Miller, French, \& Hawley 1978]{miller78} 
	Miller, J. S., French, H. B., 
	\& Hawley, S. A. 1978, \apj, 219, L85
\bibitem[Noble et al.\ 1997]{noble97} Noble, J. C., Carini, M. T.,
	 Miller, H. R., Balonek, T. J., Whitman, K., 
	\& Davis, S. M. 1997, IAU Circular 6693
\bibitem[Nesci et al.\ 1998]{nesci98} Nesci, R., Maesano, M.,
	 Massaro, E., Montagni, F., Tosti, G., 
	\& Fiorucci, M. 1998, \aap, 332, L1
\bibitem[Ohashi et al.\ 1996]{ohashi96} Ohashi, T., et al. 1996, PASJ, 48, 157
\bibitem[Padovani \& Giommi 1996]{pad96} Padovani, P., \& Giommi, P. 
	1996, MNRAS, 279, 526
\bibitem[Pian et al.\ 1998]{pian98}
	Pian, E., et al. 1998, \apj, 492, L17
\bibitem[Rybicki \& Lightman 1979]{rybicki} 
	Rybicki, G. B., \& Lightman, A. P. 1979,
	Radiative Processes in Astrophysics (New York: Wiley)
\bibitem[Sambruna, Maraschi, \& Urry 1996]{sam96} Sambruna, R. M., 
	Maraschi, L., \& Urry, C. M. 1996, \apj, 463, 444 
\bibitem[Sambruna et al.\ 1999]{rita99} Sambruna, R. M., Ghisellini, G.,
	 Hooper, E., Kollgaard, R. I.,  Pesce, J. E.,
	 \& Urry, C. M. 1999, \apj, 515, 140 
\bibitem[Sikora, Begelman, \& Rees 1994]{sikora94}
	Sikora, M., Begelman, M. C., \&Rees, M. J. 1994, \apj, 421, 153
\bibitem[Tagliaferri et al.\ 2000]{taglia00}
        Tagliaferri, G., et al. 2000, \aap, 354, 431
\bibitem[Urry et al.\ 1996]{urry96} Urry, C. M., et al. 1996, \apj, 463, 424
\bibitem[Vermeulen et al.\ 1995]{vermeulen95}
	Vermeulen, R. C., et al. 1995, \apj, 452, L5
\bibitem[Yamashita et al.\ 1997]{yamashita97}
	Yamashita, A., et al. 1997, IEEE Trans Nucl. Sci., 44, 847 

\end{thebibliography}
\end{document}